# Experimental Evidences of Quantum Phase Transition in a Spin Cluster Compound

Tanmoy Chakraborty [1], Harkirat Singh [1] and Chiranjib Mitra [1, a)]

[1]Indian Institute of Science Education and Research (IISER) Kolkata, Mohanpur Campus, PO: BCKV CampusMain Office, Mohanpur - 741246, Nadia, West Bengal, India.

a) Electronic mail: chiranjib@iiserkol.ac.in

## Abstract

Experimental realization of magnetic field induced quantum phase transition (QPT) is reported for $NH_4CuPO_4 \cdot H_2O$, a two spin cluster material with isotropic Heisenberg interaction. Experimental magnetization and specific heat data have been collected as a function of temperature and magnetic field. Experimental data have been analyzed in terms of Heisenberg dimer model. Two quantum complementary observables representing local and non-local properties of the spins are constructed using the experimental data and a clear evidence of QPT is observed through partial quantum information sharing when the magnetic field is swept through a particular value. Signature of QPT is also observed when specific heat is measured as a function of magnetic field at low temperature. Furthermore, using the experimental specific heat data, magnetic energy values are calculated and their variations are captured as a function of magnetic field and temperature.

# INTRODUCTION

Quantum phase transition (QPT) occurs in a quantum many body system when the ground state undergoes a qualitative change induced by some external tuning parameter such as magnetic field, pressure etc. [1, 2]. The system acquires a new ground state when the external control parameter is varied across a critical value. Thus, beyond this critical point, a characteristic change in the nature of the ground state occurs. QPT takes place at T=0 where there are no thermal fluctuations. Being a zero temperature phenomena, QPT is purely quantum mechanical in nature and is solely driven by quantum fluctuations. However, the existence of a quantum critical point in a many body system influences its behavior even in non-zero (sufficiently low) temperatures. Consequently, it gives an opportunity to experimentally capture QPT in real physical systems by suitably controlling the relevant parameters. There are numerous experimental evidences where QPT has been realized in solid state bulk systems such as metal, superconductors and insulators [3]. For instance, recently QPT has been observed in a magnetic insulator called $CoNb_2O_6$, a prototype of Ising spin ½ chain, when magnetic field is applied along the transverse direction of the chain axis [4]. $TlCuCl_3$ exemplifies another quantum antiferromagnet where magnetic field and pressure induced QPT have been observed and the existence of novel ordered phase is detected through inelastic neutron scattering experiments [5, 6]. Moreover, QPT has been captured in quite a few metallic compounds like ruthenium oxides [7], iron pnictides [8], copper oxides [9], heavy-fermionic systems [10] etc. In addition, quantum spin models (dimerized spin chain, spin ladder, transverse-field Ising model etc.) are very good examples of interacting many body systems where a numbers of QPT-based experiments have been successfully carried out [4-6, 12, 13]. These spin clusters are identified as perfect playgrounds for studying quantum information processing. Enormous amount of research activities have been carried out towards characterizing quantum correlations like entanglement in ground state and thermal states of these spin systems [14].

Quantum entanglement has been recognized as a necessary resource in describing numerous quantum protocols like teleportation and dense coding [15]. In early days, entanglement was believed to exist only in the atomic scale. However, in recent times, quite a few successful experiments have demonstrated that entanglement can affect the thermodynamic properties of a solid state system even at macroscopic scale [16-20]. Based on this criterion, entangled states have been detected in solid state bulk systems through well established experiments like inelastic neutron scattering, specific heat and magnetic susceptibility measurements [21]. Interestingly, it has been observed that entanglement has close connection with QPT [22-26]. Entanglement plays a crucial role in the vicinity of a QPT where correlation functions exhibit scale invariance. Osterloh *et al.* have theoretically demonstrated that entanglement shows scaling behavior close to the quantum critical point [22]. In the case of a first order QPT, the discontinuity in the first order derivative of the ground state energy is associated with a jump in the bipartite entanglement [27-29]. However, when the case of a second order QPT is considered, the critical point creates a singularity in the derivative of bipartite entanglement. The phenomenon of second order QPT has been first observed in transverse field Ising model [22] and later investigated for XY spin system [23] whereas first order QPT has been observed in frustrated spin systems [28], Heisenberg spin dimer model etc. [29, 30]. When the externally applied magnetic field is swept across a particular value, the pronounced feature of QPT can be captured in a Heisenberg spin system using a complementary relation between two experimentally measurable quantities; magnetic susceptibility and magnetization. The notion of the complementarity between these two observables was first reported by Hiesmayr *et al.* [31].

If we assume that the external magnetic field is applied along the Z-direction, then magnetization M can be defined as the sum of the expectation values of the site spins ($S_i^z$) along the direction of the applied magnetic field, i.e. $M = \sum_{i=1}^{N} \langle S_i^z \rangle$. Magnetic susceptibility has been interpreted as the sum over all two-site spin-spin correlations in a macroscopic body which can be expressed in a mathematical relationship as [32]

$$\chi = \frac{1}{K_B T} \Delta^2 (M_z)^2 \left( \sum_{i,j=1}^{N} \langle S_i^z S_j^z \rangle - \left\langle \sum_{i=1}^{N} S_i^z \right\rangle^2 \right) \qquad (1)$$

Where N is the number of spins per mole, $k_B$ is the Boltzmann constant and T is the absolute temperature. The above relation was derived assuming the fact that the Hamiltonian commutes with the Zeeman term i.e. $[H, B\sum_{i}^{N} S_i^z] = 0$ where B is the magnetic field. It has been demonstrated by Wiesniak *et al.* that the term ($\chi_x + \chi_y + \chi_z$) can serve as an entanglement witness (entanglement witness is an observable which gives the signature of entanglement in a given system [33]) where $\chi_x$, $\chi_y$ and $\chi_z$ stand for magnetic susceptibilities along x, y and z directions respectively [30]. In their entanglement detection protocol, it has been shown that the separability is governed by the inequality

$$\chi_x + \chi_y + \chi_z < \frac{1}{3} \frac{NS}{K_B T} \qquad (2)$$

It must be mentioned that in this method the full knowledge of the Hamiltonian is not necessary. Hence, magnetic susceptibility exhibits a close association with entanglement as both capture non-local properties. On the other hand, unlike susceptibility, magnetization deals with the local properties of the individual spins since it represents the average of single site spins and does not capture any two point correlation. Based on these criteria, two complementary observables were constructed; 'P', which is a function of magnetization capturing local properties and 'Q', which depends upon susceptibility capturing non-local properties. Hence, the positive value of 'Q' is an indicative of presence of entanglement [30]. For all values of the applied magnetic field, 'P' and 'Q' obey the complementarity relation $P + Q \leq 1$. In ref. [30], the above relation has been employed to capture QPT in a Heisenberg spin ½ dimer system.

The present compound under investigation is $NH_4CuPO_4 \cdot H_2O$, an antiferromagnetically coupled spin ½ Heisenberg dimer with weak interdimer interactions [34-36]. The two spin Heisenberg Hamiltonian which can well describe the present system is given by,

$$H = 2JS_1S_2 + B(S_1^z + S_2^z) \tag{3}$$

Where $J$ is the exchange coupling constant, $S_1$ and $S_2$ are the total spins at site 1 and 2 respectively, such that $S_i = S_i^x \hat{i} + S_i^y \hat{j} + S_i^z \hat{k}$. In the present paper, by means of magnetic and thermal measurements QPT is investigated in $NH_4CuPO_4 \cdot H_2O$. Field dependent isothermal magnetization measurements and temperature dependent specific heat measurements are performed on $NH_4CuPO_4 \cdot H_2O$. The experimental data are analyzed within the framework of Heisenberg isolated dimer model. Subsequently, the signature of first order QPT has been observed by charactering the entanglement properties and plateau-like behavior in magnetization data. QPT has also been captured through the field dependent specific heat data.

**EXPERIMENTS**

$NH_4CuPO_4 \cdot H_2O$ was synthesized and crystallized in single crystalline form following the synthesis route described in ref. [34]. Isothermal magnetization measurements as a function of magnetic field have been carried out in a Magnetic Property Measurement System (MPMS) by Quantum Design, USA and Vibrating Sample Magnetometer (VSM) by Oxford Instruments, UK. The temperature was varied from 2 to 10K and the magnetic field was varied from 0 to 14T. Temperature dependent specific heat measurements were performed (by standard relaxation method) in a cryogen free magnet manufactured by Cryogenic Limited, UK, at different applied magnetic field values. The temperature was varied from 2 to 10K and the field was varied from 0 to 9T. In order to get rid of the background contribution in the specific heat data, addenda measurements were performed before starting the experiments and later subtracted from the measured specific heat data.

**RESULTS AND DISCUSSIONS**

Earlier reported static magnetic susceptibility data as a function of temperature have exhibited a pronounced peak at T=6.5K which is associated with the antiferromagnetic ordering temperature [35]. The magnetic order in $NH_4CuPO_4 \cdot H_2O$ is short range in nature and arises predominantly due to intra-dimer interaction. Hence, Heisenberg dimer model was capable to fit the experimental data with an excellent agreement and yielded a value of exchange coupling constant J=5K [36]. Therefore, 2J equals the spin gap of the present compound between the singlet ground state and the tripled excited states. It is worth mentioning here that the ground state of the dimerized system can be well represented by the state $\frac{1}{\sqrt{2}}(\uparrow\downarrow - \downarrow\uparrow)$ associated with the energy eigenvalue of –3J/4 whereas the excited state is a 3-fold degenerate triplet state ($\downarrow\downarrow$, $\frac{1}{\sqrt{2}}(\uparrow\downarrow + \downarrow\uparrow)$ and $\uparrow\uparrow$) possessing energy eigenvalue of –J/4 in absence of external magnetic field [12]. The susceptibility curve suggests that one can capture the short-range antiferromagnetic correlations in only low temperatures. Therefore magnetization isotherms are taken below the ordering temperature where antiferromagnetic correlations survive significantly. Magnetic field dependent magnetization curves taken at 2K, 3K, 6K and 10K are shown in Fig. 1. The magnetic field is varied from 0T to 14T. If we focus on the magnetization isotherm at the lowest temperature (2K), we can clearly see that the magnetization almost saturates at 14T. Importantly, a step-like nature can also be observed in the magnetization curve which is a suggestive of singlet to triplet phase transition. However, the jump from one plateau to another in the magnetization curve at 2K is not sharp, rather more gradual. This is due to the fact that at finite temperature the system remains in a statistical mixture of the singlet state and three-fold degenerate triplet states. Hence, when the temperature is as low as 2K, the lowest energy state is not pure singlet state although the singlet state will have a dominant contribution in the mixture. This is the reason the jump from one plateau to another in the magnetization curve is not abrupt. However, the contribution from the singlet state goes on decreasing upon increasing temperature. Consequently, the transition between the two plateaus becomes more gradual in the isotherms taken at higher temperatures. The minimum magnetic field required to excite the system from the ground state to the first excited state is the critical field which corresponds to the excitation gap of the system. In the later sections, QPT and the identification of the critical field have also been discussed from the point of view of partial quantum information sharing and field dependent specific heat data. The experimental magnetization data have been analyzed within the framework of Heisenberg dimer model. The field dependent magnetization curves for Heisenberg dimer model have been numerically simulated for all the four isotherms and plotted on top of the experimental ones. Lande-g factor g was assumed to be 2.16 [35]. The best match between the theory and the experiment was found for J=4.9K. Subsequently, all the experimental isothermal magnetization curves were used to generate a surface plot which exhibits the distinct nature of magnetization when field and temperature both are varied. The surface plot is shown in Fig. 2.

Weakly coupled spin cluster compounds can be effectively approximated as comprising noninteracting clusters which contain a few numbers of spins, like two spins in a dimer, three spins is a trimer and so forth. $Cu(NO_3)_2 \cdot 2D_2O$ (dimer) [37], $(NHEt)_3[V_8^{IV}V_4^{IV}As_8O_{40}(H_2O)] \cdot H_2O$ (tetramar) [38], $Na_2V_3O_7$ (nine spins ring) [39] are some of the important examples of weakly coupled spin cluster materials which have been perfectly described in terms of independent clusters. Thus, the reduction in dimension reduces the dimension of the concerned Hilbert space to finite-dimensions making the theoretical calculations simpler. These spin clusters have been considered to be ideal candidates to explore QPT from quantum information theoretic point view [12, 28, 29]. For instance, Wiesniak *et al.* have illustrated the aforementioned macroscopic quantum complementarity for a cluster of two spins (dimer) and found an explicit signature of QPT [30]. The experimental validity of the complementarity is tested here for $NH_4CuPO_4 \cdot H_2O$ which also exemplifies a two spin cluster compound. The quantum complementary relation, when mathematically expressed, reads as [30],

$$\underbrace{1 - \frac{K_B T \chi}{NS}}_{Q} + \underbrace{\frac{\langle \overline{M} \rangle^2}{N^2 S^2}}_{P} \leq 1 \tag{4}$$

Where the quantity 'Q', ($Q = 1 - \frac{K_B T \chi}{NS}$) having an analytical dependence on magnetic susceptibility, accounts for the local properties of the individual spins. On the other hand, 'P' ($P = \frac{\langle \overline{M} \rangle^2}{N^2 S^2}$) depends upon magnetization and describes non-local quantum correlations between spins. In one extreme case, when the system is maximally entangled, 'Q' takes up its maximum value of one at the expense of local properties, i.e. P=0. On the other extreme, when all the individual spins of the system are aligned in the same direction, entanglement vanishes and the local properties of the spins become well defined, i.e. P=1 at the expense of 'Q'. Hence, this scenario physically signifies partial quantum information sharing between local and non-local properties of the spins. We have illustrated the complementarity relation for the case of $NH_4CuPO_4 \cdot H_2O$ and have shown that experimental magnetization and susceptibility data satisfy the inequality. The quantities 'P' and 'Q' have been constructed using the experimental data. Subsequently, 'P+Q' has been plotted in Fig. 3 as a function of magnetic field at T=2K. The plot shows a dip around B=6.9T in the 'P+Q' curve which corresponds to the quantum critical point. The theoretical values of 'P', 'Q' and 'P+Q' (at T=2K) have also been plotted with the experimental one in the same graph. It can be clearly seen from the plot that the theoretical 'P+Q' curve matches reasonably well with the experimental one. The evolution from one extreme end to another happens as the magnetic field is swept from zero to a saturation value. In absence of external magnetic field, the singlet state is the ground state and the excited states are the three-fold degenerate triplet states. When magnetic field is applied the triplet state splits into three states. At finite temperature, increase in field increases the proportion of separable triplet states and thus reduces the relative contribution of entangled states in the statistical mixture of entangled and separable states. Consequently entanglement decreases. On the other hand, as the magnetic field increases, the spins orient themselves along the applied field direction resulting in higher value of magnetization. Eventually, when magnetic field reaches a particular value, magnetization saturates. Thus, a sudden decrease in entanglement is accompanied by a sudden increase in the magnetization when the field is swept (see Fig. 3). It so happens that in the whole range of the magnetic field the relation $P + Q \leq 1$ remains valid. These abrupt changes of entanglement and magnetization with magnetic field are associated with the QPT induced by magnetic field. A surface plot of 'P+Q' as a function of magnetic field and temperature has also been generated and shown in Fig. 4. It can be seen from the plot that the dip is more pronounced at low temperature and gets broaden as the temperature increases. This is due to the fact that proportion of weight factor attributed to the singlet state in the mixture decreases at the expense of the triplet states as the temperature is increased.

Previously reported low temperature specific heat data on $NH_4CuPO_4·H_2O$ crystals have been successfully analyzed in terms of dimer model [36]. After successful subtraction of the lattice contribution from the total specific heat data (using β=0.00022K$^{-3}$ [36]), the magnetic part was efficiently fitted to isolated Heisenberg dimer model. The appearance of a rounded peak at 3.5K is the most notable feature in the specific heat curve. This is an indicative of a Schottky-like anomaly [40] which is a characteristic of a two level system and mainly occurs at low temperature due to a gradual occupation of the excited states as one varies the temperature. This observation is supported by the fact that magnetic interactions in $NH_4CuPO_4·H_2O$ can be well described by spin cluster model. Fig. 5 exhibits temperature dependent molar specific heat curves (lattice part subtracted) of the crystals in zero field and in different externally applied magnetic fields. Temperature is varied from 2K to 10K and the field is varied from 0T to 9T. It can be clearly observed from the plot that the Schottky-like peak is significantly affected by the magnetic fields. A remarkable lowering and simultaneous broadening in the maxima happens when the field is swept from lower value to upper values. This can be qualitatively understood in the framework of the energy splitting as a function of field as shown in Fig. 6. This distinct feature in the specific heat data is well consistent with theoretical predictions made by Bonner and Fisher [41]. Furthermore, in order to reveal the behavior of specific heat as a function of magnetic field and temperature, a surface plot is created and depicted in Fig. 7. The plot provides a pictorial representation of the evolution of the specific heat with magnetic field as the temperature of the system changes. This scenario has been discussed in more details in the subsequent section.

In this section of the paper, the variation of specific heat has been explored as a function of magnetic field and an evidence of QPT has been witnessed. Externally applied magnetic field causes elementary excitations and changes the energies of the eigenstates of an antiferromagnetically dimerized system [41, 42]. Hence, when the field is swept across a particular value, a level crossing occurs between the ground state and the first excited state [12]. Interestingly, such a transition from one energy state to the other influences certain physical properties like specific heat at low temperatures. Isothermal specific heat data (at 2K) as a function of magnetic field is exhibited in Fig. 8. Below the critical field value, the system remains in the singlet ground state corresponding to the energy $-\frac{3J}{4}$. The first excited state is one of the triplet states with the associated energy $\frac{J}{4} - g\mu_B B$. On increasing the magnetic field, the first excited state approaches the ground state and the excitation gap reduces as shown in the Fig. 6. At the quantum critical point, the ground state and the excited states possess the same energy and the spins can be excited from ground state to the first excited state with a minimal thermal energy which is responsible for a dip (at 6.9T) in the field dependent specific heat curve (Fig. 8). Beyond this level crossing point, the erstwhile first excited state becomes the new ground state. The energy gap between the ground sate and the first excited state goes on increasing monotonically as magnetic field increases further. As the temperature increases, the dip becomes less pronounced and dies out at high temperatures. One can understand this feature as follows. The above mentioned level crossing scenario occurs in the ideal zero temperature case when the system is in a pure state. However, at finite temperature the system is in a mixture of the all four states. If the ground state is considered, the proportion of the singlet state in the mixture decreases with increasing temperature. Consequently, the change in internal energy as a function of field becomes less sharp at QCP. As a direct consequence, the dip in the specific heat curve also becomes less pronounced (as the specific heat is connected with internal energy through simple mathematical relationship). Fig. 9 shows field dependent specific heat curves at temperatures 2.5K, 4K, 5K and 7K. The gradually broadening nature of the dip is clearly evident from these plots. Thus, these arguments establish the fact that the appearance of the dip in the isothermal specific heat curve is solely due to the level crossing driven by externally applied magnetic field. Heisenberg dimer model has been employed to analyze the field dependent specific heat isotherm. The specific heat can be calculated theoretically using the following equation.

$$C_v = \frac{1}{K_B T^2}(\langle H^2 \rangle - \langle H \rangle^2) \tag{5}$$

Where $K_B$ is the Boltzmann constant and H is the Heisenberg dimer Hamiltonian. The variation of specific heat with field has been simulated using the above equation (substituting J=4.9K) and plotted on top of the experimental data in Fig. 8. One can conclude that the simulated curve is in good agreement with the experimental data.

Herein, using the experimental specific heat data internal energy is estimated and its variation is captured as a function of temperature and magnetic field. At some particular temperature T, the internal energy U(T) can be expressed in terms of specific heat $C_p(T)$ by the mathematical relation given as

$$U(T) = U_2 + \int_2^T C_v(T) dT \qquad (6)$$

Where $U_2$ being the internal energy at 2K. The magnetic part of the temperature dependent specific heat data were integrated numerically and substituted in the above equation. Thus internal energy is quantified for $NH_4CuPO_4 \cdot H_2O$ as a function of temperature. The above analysis was performed for all the field dependent datasets. The proper variation of the integration constants ($U_2$) with field are determined theoretically and incorporated in the integrations. Both the theoretical and the experimental energies are scaled in units of Kelvin. Quantified internal energies as a function of temperature for different applied magnetic fields are plotted in the same graph and are shown in Fig. 10. These U(T) vs. T datasets are used to generate a surface plot (Fig. 11) which explicitly depicts the behavior of internal energy when both the magnetic field and the temperature are varied. The plot clearly shows that the internal energy as a function of field becomes less sharp as the temperature increases which supports our previous discussion. With a motivation to compare the experimental plot with theoretical prediction, the surface plot of internal energy as a function of temperature and magnetic field is generated for Heisenberg spin ½ dimer model and exhibited in Fig. 12. One can conclude that these two plots are remarkably consistent with each other which points towards successful experimental quantification of magnetic energy for $NH_4CuPO_4 \cdot H_2O$.

**CONCLUSIONS**

The present work is an example where QPT has been captured in a physical system by carrying out experimental measurements of macroscopic thermodynamic properties like magnetization and specific heat in the thermodynamic limit. Signature of QPT is explored in $NH_4CuPO_4 \cdot H_2O$ from different angles which can be summarized as follows. A step like nature has been observed in the field dependent magnetization data which coexists with the plateaus in the entanglement curve, indicating the characteristic of first order quantum phase transition occurring due to crossing of the energy levels. Quantum complementarity relation between two observables representing local and nonlocal properties has been experimentally verified using the experimental data. Subsequently, we have examined the behavior of experimental specific heat data as a function of magnetic field and captured the critical point which corresponds to a dip in the specific heat curve. It has been established that QPT occurs solely due to the level crossing between the ground state and the first excited state induced by external magnetic field which leads to the appearance of the dip in the field dependent specific heat curve. Moreover, in order to gain additional details about the system, magnetic energy has been extracted and its variations are investigated as a function of temperature and magnetic field. Conclusively, field dependent magnetization and specific heat data have shown excellent consistency in capturing the quantum critical behavior of $NH_4CuPO_4 \cdot H_2O$.

**ACKNOWLEDGMENT**

The authors would like to thank the Ministry of Human Resource and Development, Government of India for funding. The authors are grateful to Prof. S. K. Dhar for allowing us to carry out high field magnetic measurements in his lab in TIFR, Mumbai. We also acknowledge Arvind Maurya for his assistance with the magnetic measurements. The authors also would like thank to Dr. Swadhin Mandal for allowing us to use his lab facilities for the synthesis of the system.

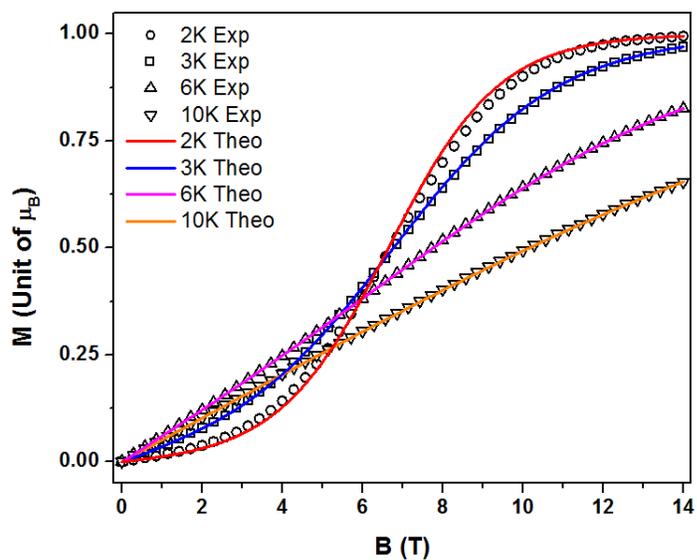

FIG. 1. Experimentally measured magnetization vs. magnetic field at different temperatures (as shown in the legend) along with the corresponding simulated curves derived using the Hamiltonian for Heisenberg dimer.

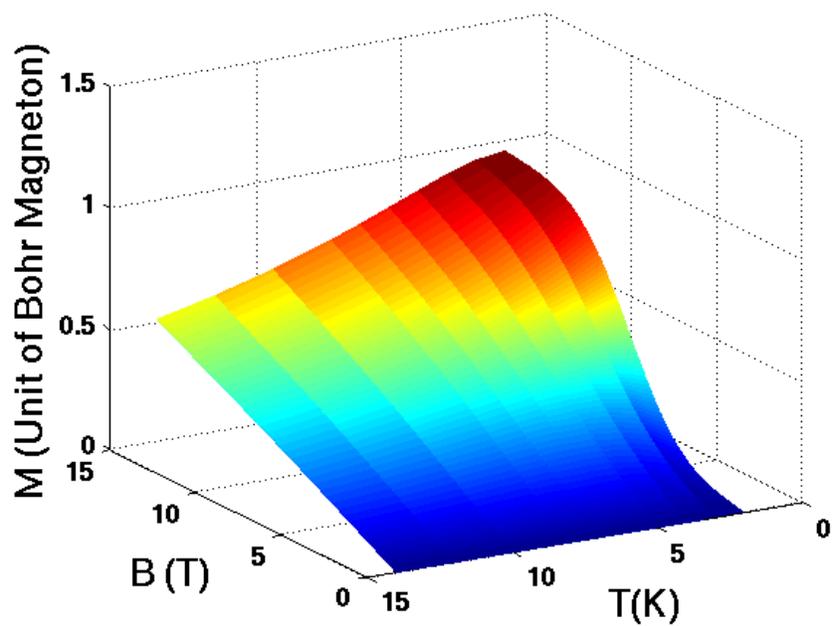

FIG. 2. Surface plot with magnetization, magnetic field and temperature along the three axes.

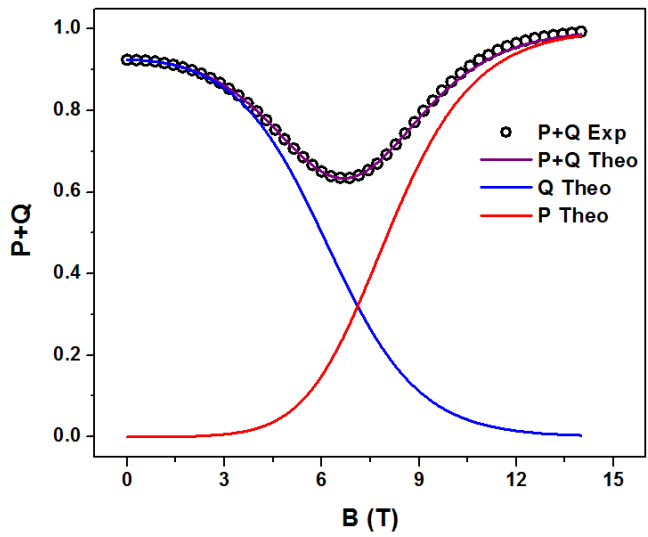

FIG. 3. Plot of 'P', 'Q' and 'P+Q' as function of magnetic field.

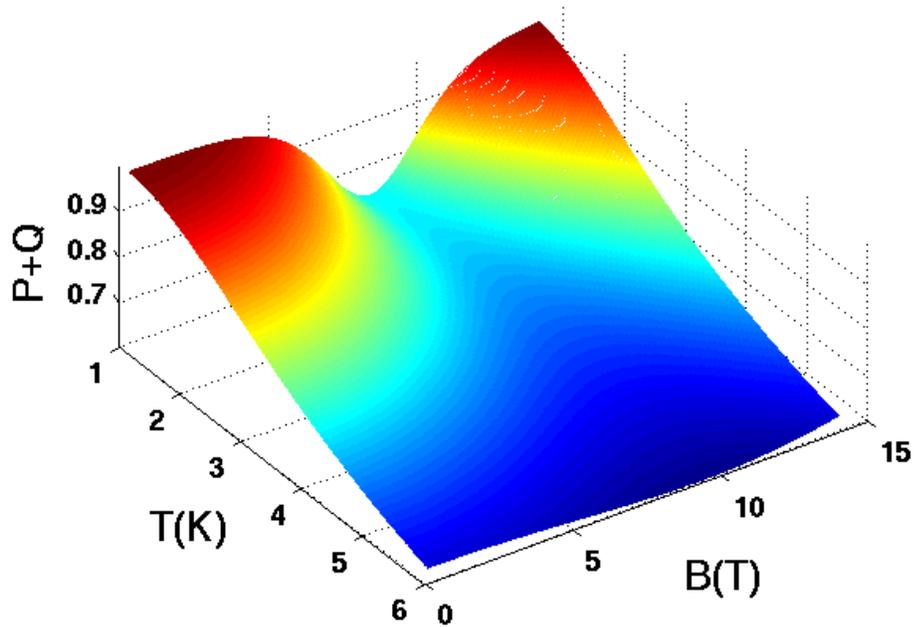

FIG. 4. Surface plot of 'P+Q' as a function of temperature and magnetic field.

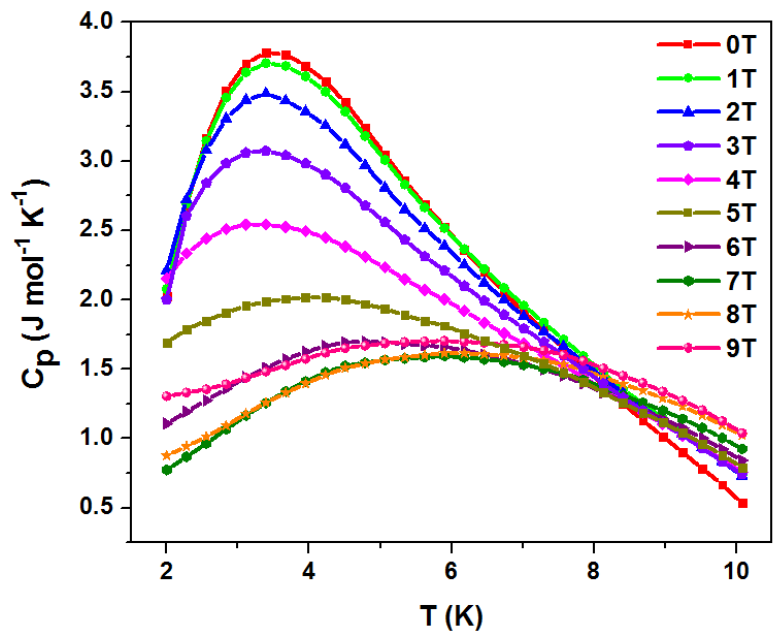

FIG. 5. Experimental specific heat data for $NH_4CuPO_4 \cdot H_2O$ as a function of temperature at different fields as shown in the legend.

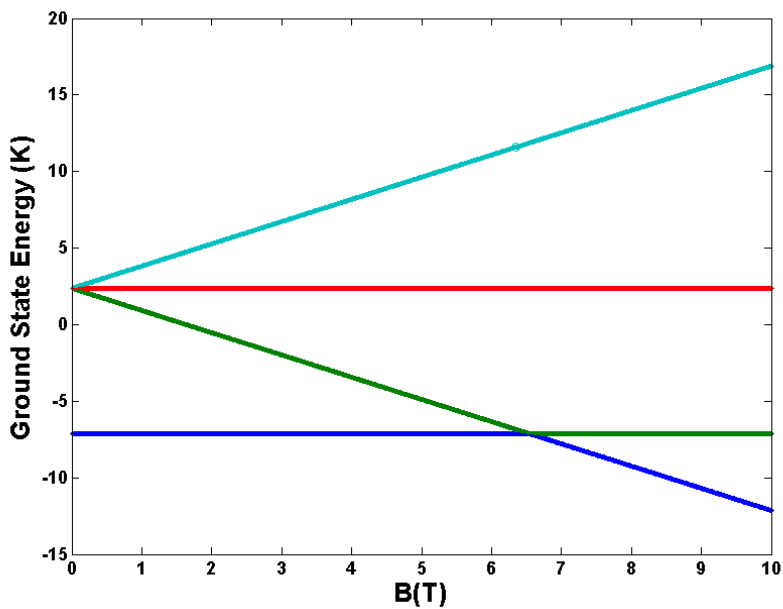

FIG. 6. Energy eigenvalues versus magnetic field for a dimer system with J=4.9K.

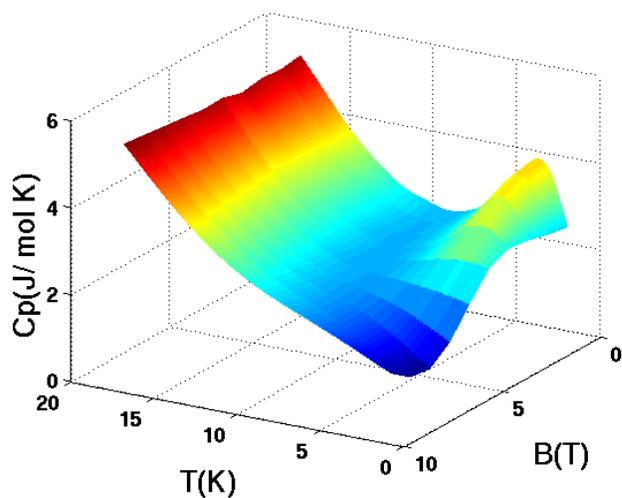

FIG. 7. Three dimensional plot depicting the variation of experimental specific heat with magnetic field and temperature.

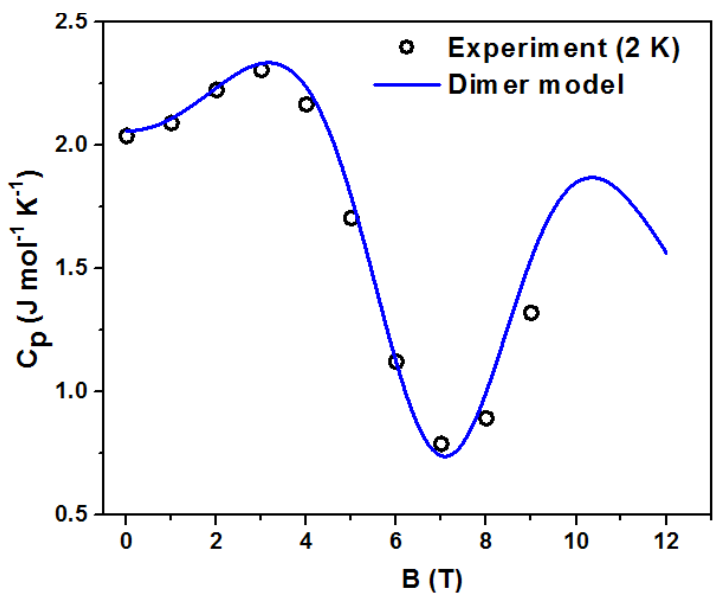

FIG. 8. Isothermal specific heat for $NH_4CuPO_4 \cdot H_2O$ as a function of magnetic field (at 2K). Circles represent the experimental data and the solid red line is the theoretical prediction simulated numerically for Heisenberg dimer model.

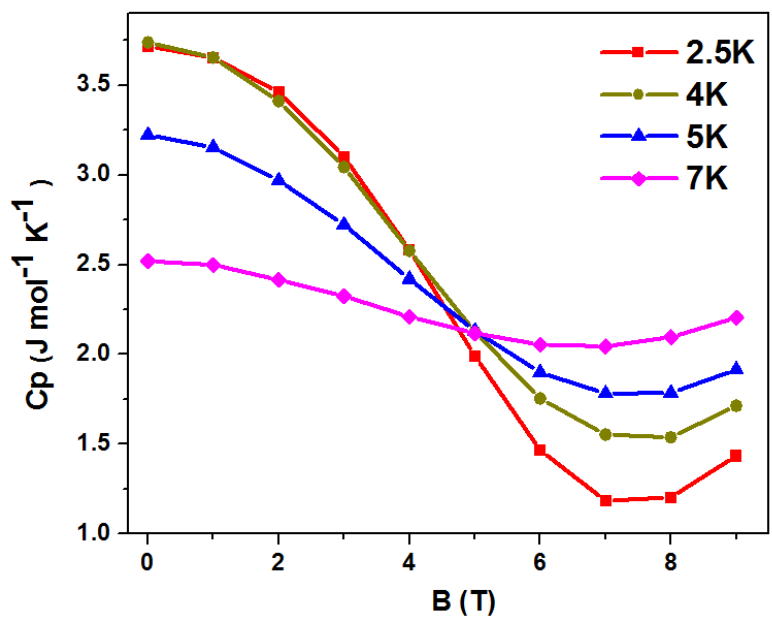

FIG. 9. Experimental $C_p$ vs. B data at different temperatures as mentioned in the legend.

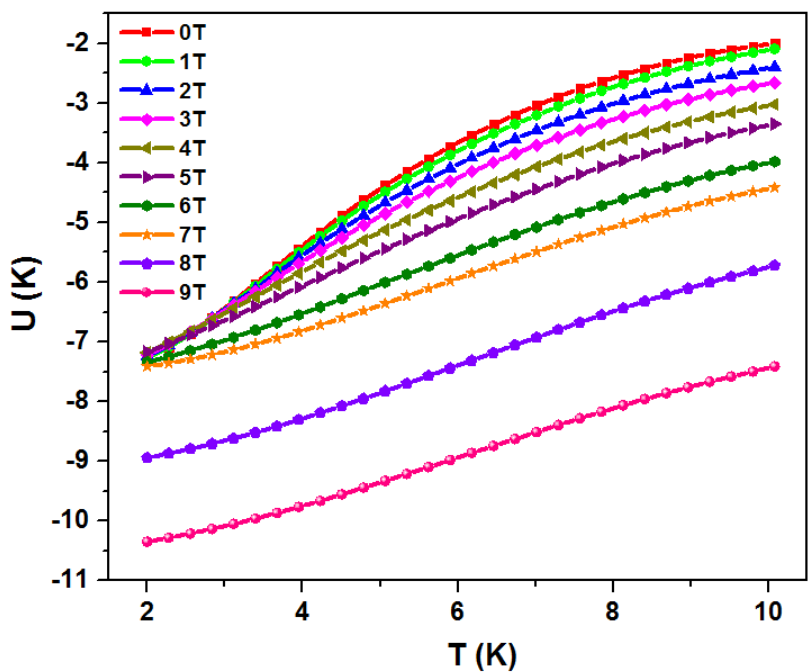

FIG. 10. Variation of experimental Internal energy with temperature for different applied magnetic fields.

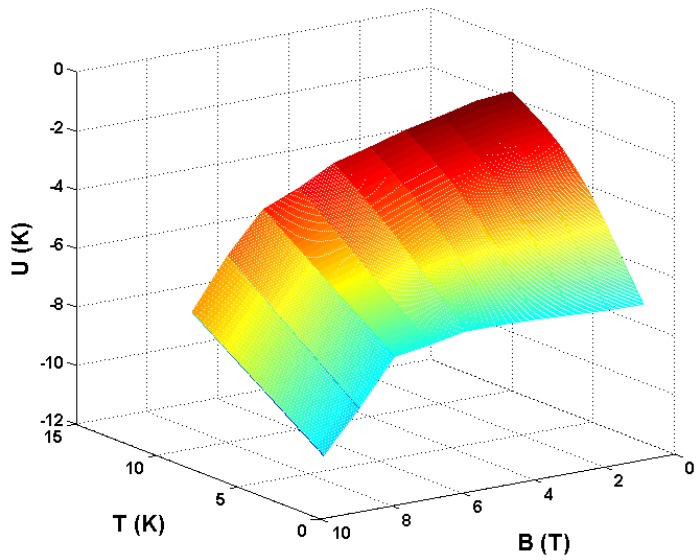

FIG. 11. Three dimensional variation of experimental internal energy with magnetic field and temperature along the other two axes.

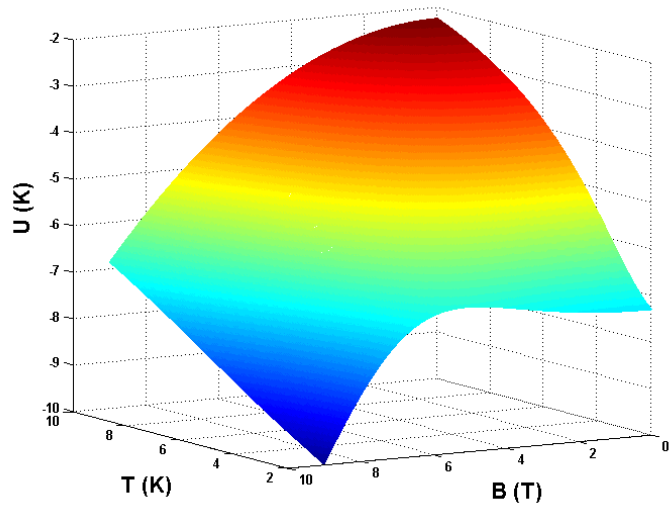

FIG. 12. Surface plot depicting internal energy as a function of temperature and magnetic field for Heisenberg spin ½ dimer model.